\documentclass{aastex}          %% The manuscript based on AASTeX v5.x
\usepackage{spr-astr-addons}    %% mimicing ASTR journal style
\usepackage{soul}
\usepackage{epsfig, graphicx, amsmath, amsfonts, amssymb, mathrsfs, textcomp, array}
\usepackage{bm}
\def\st#1{}

\begin{document}
%% Article title
%
\title{The adaptive-loop-gain adaptive-scale CLEAN deconvolution of radio interferometric images}

%% Running heads
%\shorttitle{<Short article title>}
%\shortauthors{<Autors et al.>}

%% Author and Affilations
\author{L. Zhang\altaffilmark{1,3}}
%\affil{Xinjiang Astronomical Observatory, Chinese Academy of Sciences, 150 Science 1-Street, Urumqi 830011, P.R. China}
\email{lizhang.science@gmail.com}

\and

\author{M. Zhang\altaffilmark{1,2}}
%\affil{University of Chinese Academy of Sciences, Beijing 100049, P.R. China}
%\email{zhangming@xao.ac.cn}

\and

\author{X. Liu\altaffilmark{1,2}}
%\affil{Key Laboratory of Radio Astronomy, Chinese Academy of Sciences, Nanjing 210008, P.R. China}
%\email{} %% non-output
%\email{liux@xao.ac.cn}

%% Alternate Affilations
\altaffiltext{1}{Xinjiang Astronomical Observatory, Chinese Academy of Sciences, Urumqi 830011, P.R. China}\email{lizhang.science@gmail.com}
\altaffiltext{2}{Key Laboratory of Radio Astronomy, Chinese Academy of Sciences, Urumqi 830011, P.R. China}
\altaffiltext{3}{University of Chinese Academy of Sciences, Beijing 100049, P.R. China}

%% Abstract
\begin{abstract}
  CLEAN algorithms are a class of deconvolution solvers which are
  widely used to remove the effect of the telescope
  Point Spread Function (PSF). Loop gain is one
  important parameter in CLEAN algorithms. Currently the parameter is fixed during
  deconvolution, which restricts the performance of CLEAN
  algorithms. In this paper, we propose a new deconvolution algorithm
  with an adaptive loop gain scheme, which is referred to as the
  adaptive-loop-gain adaptive-scale CLEAN (Algas-Clean) algorithm. The
  test results show that the new algorithm can give a more accurate
  model with faster convergence.
\end{abstract}

%% Keywords
\keywords{methods: data analysis; techniques: image processing}

\section{Introduction}

Aperture synthesis technique breaks through the limitation of
resolution of a single antenna physical aperture. Such
telescopes however do not measure all the spatial frequencies, leading
to a Point Spread Function (PSF) with wide-spread sidelobes. The PSF
limits the imaging dynamic range to only a few 100:1.

There are many methods to remove the effects of the PSF, e.g. CLEAN
algorithms \citep{hog74, cla80, sch83, bha04, cor08, rau11}, Maximum
Entropy Methods (MEM) \citep{cor85,nar86} and compressive sensing
reconstruction algorithms \citep{wia09, wen10, li11, car12}. The CLEAN algorithm \citep{hog74} and its variants model the sky
  brightness distribution as a set of delta functions.  This however
  is non-optimal to model extended emission. To improve the
imaging performance for extended emission, several
  scale-sensitive algorithms have been proposed \citep{ bha04, cor08, rau11}. The multi-scale CLEAN algorithms \citep{cor08, rau11}
use tapered paraboloids to decompose sky sources by a
matched-filtering technique. Better representation for extended
sources makes the multi-scale CLEAN algorithms be able to use a
larger loop gain, e.g. $0.5$ or even larger \citep{cor08}. However it uses a fixed set of components, the size of which cannot
  be varied. The adaptive scale pixel decomposition deconvolution
  (Asp-Clean) algorithm \citep{ bha04} uses an optimization technique
  to overcome this problem by keeping the size and location of the
  components variable. All these algorithms use the
  loop gain parameter which controls the feed-back of the model in
  computing the residuals at each iteration. The value of the
  loop gain therefore strongly impacts both the imaging performance
  and rate of convergence.  The value of loop gain is fixed, which
  limits the performance of these algorithms. In this paper
therefore, we propose an adaptive loop gain scheme to improve the
fidelity of model and convergence speed of the Asp-Clean algorithm.

In section 2, we recap CLEAN algorithms. In section 3, we describe the
motivation of developing an adaptive loop gain scheme. In section 4,
we describe the details of the Algas-Clean algorithm. In section 5,
some examples are provided to show the performance of the Algas-Clean
algorithm. In section 6, we summary this work.

\section{CLEAN algorithms}

Interferometric measurement is in spatial frequency domain,
\begin{equation}\label{1}
\bm{V}^{measured}= \bm{S}\left( \bm{F}\bm{I}^{true} + \bm{n}_{0} \right),
\end{equation}
where $\bm{V}^{measured}$ is the measured visibility data, $\bm{S}$ is
the sampling function which encodes the missing spatial frequency
  information, $\bm{F}$ is the Fourier transformation,
$\bm{I}^{true}$ is the true sky image and $\bm{n}_{0}$ is the random
noise in the visibility domain \citep{tho01}. In the image plane, the above
equation is
\begin{equation}\label{2}
\bm{I}^{dirty}= \bm{B}\ast \bm{I}^{true}+\bm{n},
\end{equation}
where $\bm{I}^{dirty}$ is the dirty image which is the inverse Fourier
transformation of $\bm{V}^{measured}$, $\bm{B}$ is the dirty
beam which is the inverse Fourier transformation of $\bm{S}$, the symbol $\ast$ denotes the convolution, and $\bm{n}$ is
the measurement noise.

Various deconvolution algorithms exist to remove the effects of $\bm{B}$
  from $\bm{I}^{dirty}$.  Deconvolution algorithms are fundamentally
  iterative and all modern CLEAN deconvolution algorithms have the general
  structure of two iterative cycles called ``major cycle'' and ``minor
  cycle'' \citep{cla80, rau09}.  The major cycle involves computation of the residual image
  at each iteration while the minor cycle involves deconvolution of
  $\bm{B}$ (or it's approximation) from the residual image at each
  iteration. Scale-insensitive CLEAN algorithms parameterize the
sky image as a set of delta functions. The components are estimated by
finding the brightest peak from the current residual image and scaling
the peak with a fixed loop gain, $g$. Empirically, $g$ is typically
in the range of $0.01 \sim 0.25$ \citep{tay99}. Scale-sensitive algorithms like Asp-Clean and MS-Clean parameterize the sky image
with scale basis functions. The main
difference is that the Asp-Clean algorithm parameterizes the sky
brightness function with a continuous set of scales while the MS-Clean
algorithm uses several discrete scales. These scale basis functions represent the extended
structures much better than delta functions. However, since the components
cannot always accurately model the complex extended emission, loop
gain is still needed in these algorithms.  Existing implementations
use a fixed loop gain.

\section{The motivation for an adaptive scheme for loop gains}

In scale-sensitive CLEAN algorithms, a component is calculated as a scale basis function from
the region with the highest {\it total} power in the current residual image. A fixed loop gain, which scales the component before subtraction, may be proper for some components but not for
others. What's more, a fixed loop gain gives
each pixel of the \textit{extended component} equal loop gains in the
Asp-Clean and MS-Clean algorithms. This means that strong brightness
and weak brightness in \textit{an extended component} which are
estimated from the current residual image are deemed to have same
  significance/reliability. This will be a problem when a tailed
function like Gaussian is used, like in the Asp-Clean algorithm, to
approximate a finite-support extended structure. Variable
loop gain for strong brightness and weak brightness can be a better
strategy. In this paper, we propose an scheme for adaptive loop
gains. The basic idea is that stronger brightness owns larger loop gains than weaker brightness in {\it an extended component}. Obviously, the adaptive loop gains are
shape-dependent. This can solve the problem just mentioned. From lots
of tests, we found that the largest remaining errors in an image model
are in the strongest brightness region. The over-estimation of the
strong brightness leads to this problem. This problem can be solved by
suppressing the strong part of each component. To sum up, we want to {\it relatively} suppress the strongest and weakest parts of an initial component (before subtraction) to solve the two problems mentioned above.

\section{The adaptive-loop-gain adaptive-scale CLEAN algorithm}

As mentioned above, the Algas-Clean algorithm is a variant of
the Asp-Clean algorithm with the proposed adaptive loop gain scheme
used in the minor cycle.

The Asp-Clean algorithm represents images as a set of truncated Gaussian functions,
\begin{equation}\label{3}
\bm{I}^{model}= \sum_{i=1}^{N} g \bm{I}_{i}^{comp}\left(scale \right),
\end{equation}
where $\bm{I}^{model}$ is the model composed of $N$ components, $g$ is
the fixed loop gain which is used for all components, and $I_{i}^{comp} \left(scale \right)$ is the
$i$th component which is a truncated Gaussian function. The Gaussian
components are found by an optimization method like
Levenberg-Marquardt method \citep{mar63} which minimizes the objective
function $\chi^{2}$,
\begin{equation}\label{4}
\chi^{2} = \| \bm{I}_{i}^{residual} - \bm{B} \ast \bm{I}_{i}^{comp} \|_{2}^{2},
\end{equation}
where $\bm{I}_{i}^{residual}$ is the residual
image in $i$th iteration, and $\| \,\, \|_{2}$ is the $l2$ norm. The optimization method makes components much
better match to source structures and so the Asp-Clean algorithm
  gives better imaging performance compared to other algorithms. But
local and specific structures are almost always non-Gaussians, and in fact not fitted exactly by any function in general. The fixed
loop gain is not optimal for an extended component, especially for the two
problems mentioned in section 3 and can lead to an improper
approximation of components. The resulting error is difficult to be corrected completely. With
the proposed adaptive loop gain scheme, the approximated model is
expressed as follows,
\begin{equation}\label{5}
\bm{I}^{model} =  \sum_{i=1}^{N} \bm{g}_{i} \bm{I}_{i}^{comp},
\end{equation}
where $\bm{g}_{i}$ is a matrix whose elements are proportional to each pixel amplitude of the $i$th component.

A fixed loop gain equally treats strong brightness and weak
brightness. This often leads to an over-estimation for weak brightness
in finite-support case. In the Algas-Clean algorithm, each
Gaussian component will be scaled with adaptive loop gains. It can
effectively control the over-estimation of weak brightness in
finite-support structures. To reduce the computational complexity,
previous components will be not re-optimized in later iterations in
the Algas-Clean algorithm as well as in the Asp-Clean algorithm
implemented in this paper. Deconvolution is terminated when the
standard deviation of residual image is less than that of noise
included in the dirty image.

In this paper, we use the following method to calculate the adaptive loop gains,
\begin{equation}\label{6}
g\left( x \right)= \sqrt{\left |a \left( 1- f \left( x \right) \right) + b f\left( x \right) \right|},
\end{equation}
where
\begin{equation}\label{7}
f\left( x \right)= \frac{x-C_{min}}{C_{max}-C_{min}},
\end{equation}
where the $g \left( x \right)$ is the adaptive loop gain function, $x$
is an image pixel, $a$ and $b$ are the minimum and maximum values
respectively after the linear transformation, $C_{min}$ and $C_{max}$
are the minimum and maximum values of the component respectively,
$\sqrt{\,\,}$ is a root square operation and $| \,\, |$ is an absolute
value sign. The parameters $a$ and $b$ need to be assigned by users and $a \geq 0.0$ and $b \leq 1.0$. As mentioned earlier, the largest remaining errors in an image model are in the strongest brightness region. The square root operation in the formula \eqref{6} can {\it relatively} suppress both the weakest and strongest parts of an extended component. Suppression of the strongest parts helps in reducing the residual error while suppression of the weakest parts helps in limiting the error due to the tail of the
Gaussian components.

\section{Numerical experiment}

\begin{figure}[htbp]
\vspace*{0.2mm}
\begin{center}
\includegraphics[width=7cm]{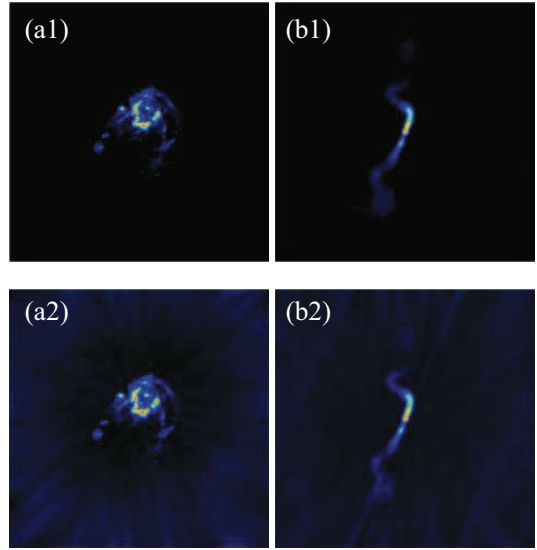}
\end{center}
\caption{The testing data are shown here. The first and second rows are the original images and the dirty images respectively. The M31 image is shown in the first column, and the 3C31 image is shown in the second column. The brightness ranges of the two original images are from $0$ \textrm{Jy/pixel} to $0.1$ \textrm{Jy/pixel}. The brightness ranges of the dirty M31 image and the dirty 3C31 image are from $-0.040$ \textrm{Jy/pixel} to $0.790$ \textrm{Jy/pixel} and from $-0.036$ \textrm{Jy/pixel} to $0.839$ \textrm{Jy/pixel} respectively. }
\label{Test-Images}
\end{figure}

To show the performance of the Algas-Clean algorithm, two simulated VLA B configuration were made with the CASA
\footnote{This is a radio astronomical data processing software
  developed by the National Radio Astronomy Observatory (NRAO). Its
  homepage is http://casa.nrao.edu/} software. The two test images M31
and 3C31 are used as sky models and are displayed in Fig~\ref{Test-Images}. The original images are available
from the NRAO's websites \footnote{The original M31 image is available
  from http://www.cv.nrao.edu/~awootten/mmaimcal/ImLib.html and the
  original 3C31 image is available from
  http://www.cv.nrao.edu/~abridle/3c31.html}. Background features
  in the images were removed with a threshold, the images are scaled to the brightness
  range from 0 \textrm{Jy} to 0.1 \textrm{Jy} and padded with zeros to
  make the size $512 \times 512$ \textrm{pixels}. The total fluxes of
  the M31 and 3C31 images are 148.584 \textrm{Jy} and 88.453
  \textrm{Jy} respectively. Gaussian noise was added to the
  simulated visibilities and robust weighting scheme \citep{bri95} (with the CASA parameters ``weighting''=``briggs'' and ``robust''=0) used during
  the imaging process.

The Root of Mean Squares (RMS) of the model error image, which is the difference between the true image and the model image, is used
  to compare the imaging performance of reconstructions.  The $RMS_{Merr}$ of a model error image is defined as
\begin{equation}\label{8}
RMS_{Merr}=\frac{\| \bm{I}^{Merr} \|_{2}}{\sqrt{M}},
\end{equation}
where $\bm{I}^{Merr} = \bm{I}^{true} - \bm{I}^{model}$ is the
model error image and $M$ is the number of pixels in the model error
image. Fewer and weaker
structures in a model error image means that the model is closer to
the underlying true image, and therefore a better reconstruction.

In the Asp-Clean algorithm, after a component is calculated by an
optimization method, the component will be scaled by a loop gain
  for all pixels of the component. In the Algas-Clean algorithm, loop gains are
adaptive for each pixel of each component. For Gaussian components,
the adaptive loop gain scheme will give large loop gains to the
  pixels near the center of the Gaussian and small loop gains for the wing region far from the
center. The wing regions of a Gaussian function make
components have a very large support and therefore non-optimal to represent
  structures with finite support. In the Algas-Clean algorithm, the
adaptive loop gain scheme will scale these values of components in the wings of the Gaussian with small loop
gains. The adaptive loop gains can be truncated when they are less
than a threshold. For example, a loop gain that is less than a factor of 0.001 of the
maximum loop gain will be set as zero. This
  effectively makes Asp-Clean components tapered with better
  support size with the added advantage that the tapering adapts to
  the signal strength.

In the test, we include results from the Clark-Clean algorithm for completeness
and to also motivate the fact that for modelling the extended
emission, scale-sensitive algorithms like Asp-Clean give much better
performance and hence we chose to optimize the Asp-Clean algorithm.
A fixed loop gain of $0.1$ is used for the Clark-Clean
algorithm and $0.7$ for the Asp-Clean
algorithm. The adaptive loop gains of the Algas-Clean algorithm vary in
the range from $0$ to $0.7$.

\begin{figure*}[htbp]
\vspace*{0.2mm}
\begin{center}
\includegraphics[width=16cm]{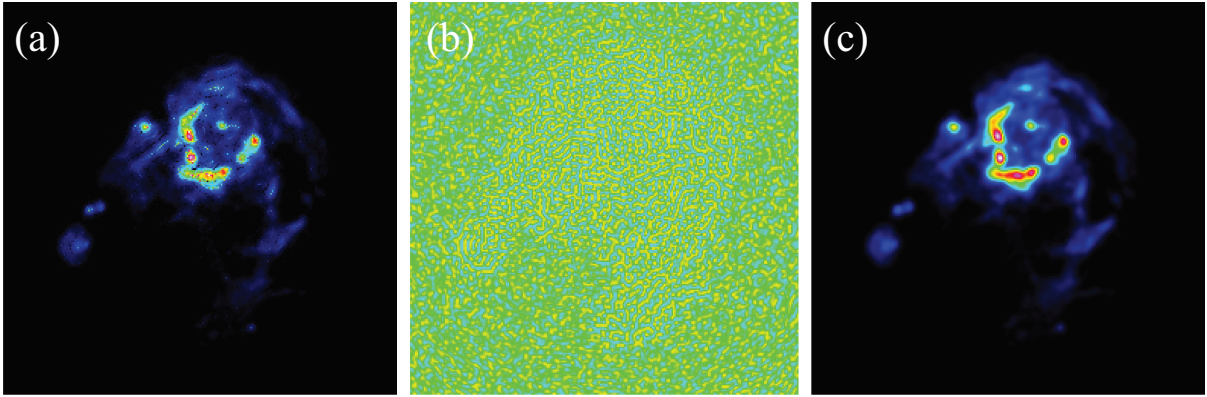}
\end{center}
\caption{The figure shows simulated results of source M31, from the Algas-Clean algorithm. (a) The model image is displayed with the brightness range from $-0.001$ \textrm{Jy/pixel} to $0.103$ \textrm{Jy/pixel}. (b) The residual image is displayed with the brightness range from $-0.0005$ \textrm{Jy/pixel} to $0.0005$ \textrm{Jy/pixel}. The restored image is displayed with the brightness range from $-0.001$ \textrm{Jy/pixel} to $0.858$ \textrm{Jy/pixel}. }
\label{FIG-M31}
\end{figure*}

To show that as with the Asp-Clean algorithm, the Algas-Clean algorithm also models
  the extended emission well and fundamentally separates signal from
  noise leaving noise-like residuals, we show the deconvolution of the
  M31 images in Fig~\ref{FIG-M31}. The total flux of
the model image in Fig~\ref{FIG-M31}(a) is $148.594$ \textrm{Jy}. The residuals
are uncorrelated and noise-like implying good imaging performance of the
Algas-Clean algorithm.

\begin{figure*}[htbp]
\vspace*{0.2mm}
\begin{center}
\includegraphics[width=16cm]{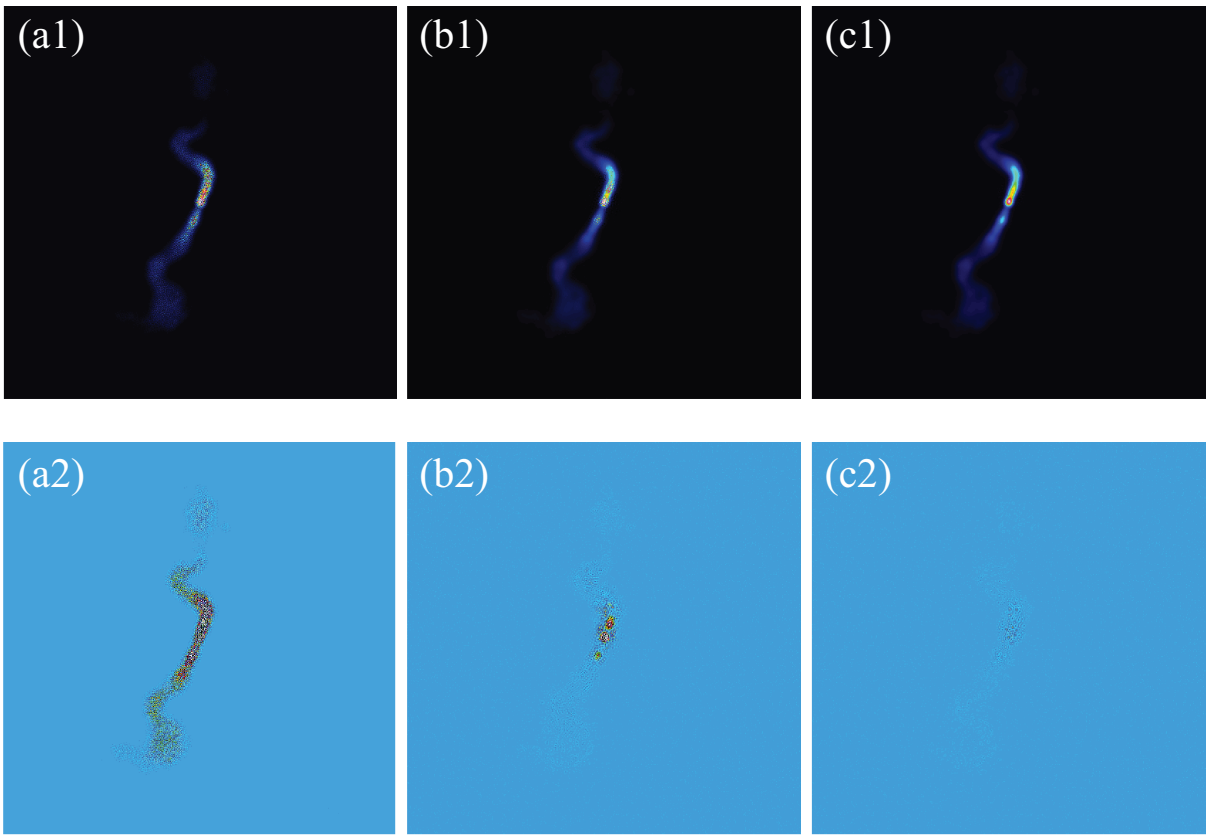}
\end{center}
\caption{The figure shows simulated results of source M31. The first, second and third columns are from the Clark-Clean, the Asp-Clean algorithm and the Algas-Clean algorithm respectively. The model images are displayed in the first row with the brightness range from $-0.001$ \textrm{Jy/pixel} to $0.1$ \textrm{Jy/pixel}. The model error images are displayed in the second row with the brightness range from $-0.01$ \textrm{Jy/pixel} to $0.02$ \textrm{Jy/pixel}. }
\label{FIG-3C31}
\end{figure*}

The test results of the 3C31 image are displayed in Fig~\ref{FIG-3C31}. We compare
the results from the widely used scale-insensitive
Clark-Clean algorithm, scale-sensitive Asp-Clean
algorithm and the Algas-Clean algorithm. Table \ref{Table-1} shows the total reconstructed fluxes of $88.253$ \textrm{Jy}, $88.472$
\textrm{Jy} and $88.481$ \textrm{Jy} for the Clark-Clean, the
Asp-Clean algorithm and the Algas-Clean algorithm respectively. They
are very close to the total flux of the true image. Comparing the results of the Clark-Clean algorithm and the
Asp-Clean algorithm, we found that the scale-sensitive CLEAN algorithm
has significantly improved the fidelity of model for extended
sources. However, you can see that there are still lots of the
remaining errors in the Asp-Clean model. The two problems mentioned
earlier are the main reasons for the remaining errors in the Asp-Clean
model. With the help of the proposed loop gain scheme, the Algas-Clean
algorithm fundamentally solves the two problems. From Table \ref{Table-1}, you can
see that the RMS level of the remaining errors is reduced by a order
of magnitude. This can also be verified by comparing Fig~\ref{FIG-3C31}(b2) and
Fig~\ref{FIG-3C31}(c2).

%%%%%%%%%%%%%%%%%%%%%%%%%%%%%%%%%%%%%%%%%%%table 1 numerical comparison %%%%%%%%%%%%%%%%%%%%%%%%%%%%%%%%%%%
%\clearpage
 %\onecolumn
  \begin{table*}
   \caption{The numerical comparison when the 3C31 residuals are in the noise level}
  %{\label{kstars} Table 1. Source Positions}
    \label{table:1}
     \centering
\begin{tabular}
%{p{3.3cm}p{1.7cm}p{1.7cm}p{1.7cm}}
{c c c c c c c}
\hline\hline % inserts double horizontal lines
 & Gaussian & Delta & $RMS_{Merr}$ & Total model flux\\
 & components & components & \textrm{(Jy/pixel)} & \textrm{(Jy)} \\
\hline
Clark  & & $104000$ & $1.584 \times 10^{-3}$  & $88.253$\\
\hline
Asp & $1400$ & $48800$ & $7.795 \times 10^{-4}$  & $88.472$\\
\hline
Algas & $1500$ & $17200$ & $7.623 \times 10^{-5}$  & $88.481$\\
\hline
%Asp-Clean (G21)&  & & & &\\
%\hline
%Algas-Clean (G21)& &  & & &\\
%\hline
 \end{tabular}
  %\tablecomments{RTF denotes the Reconstructed Total Flux.}
   \label{Table-1}
   \end{table*}

   From Table \ref{Table-1}, we also see that the Algas-Clean model is composed of
   Gaussian components and delta
     functions. These compact (delta) components are from small
   residuals whose widths are smaller than the
   width of the main lobe of the dirty beam. We also can see that fewer compact components are included in
   the model image of the Algas-Clean algorithm indicating that
     Algas-Clean models the extended emission better than Asp-Clean
     algorithm. This means that the decomposition is more effective,
   and fewer number of iterations are needed in the Algas-Clean
   algorithm. Since the Asp-Clean algorithm is compute-intensive, fewer
   iterations will significantly speed it up. From our tests, we have
   found that the runtime of the Algas-Clean algorithm is about half the runtime of the Asp-Clean algorithm -- i.e.,
     Algas-Clean is $\sim 50$\% faster than Asp-Clean with a better
     imaging performance. In a future paper, we will report on
     optimizing the fitting procedure used in Asp-Clean to further
     improves its run-time performance.

   Because the scale-sensitive CLEAN algorithms parameterize the true
   image in a
     scale-sensitive basis, it is obvious that the proposed adaptive
   loop gain scheme will have the similar performance as the
   Algas-Clean algorithm for other scale-sensitive CLEAN algorithms.

\begin{figure}[htbp]
\vspace*{0.2mm}
\begin{center}
\includegraphics[width=8cm]{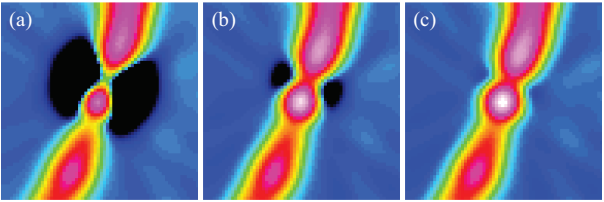}
\end{center}
\caption{The figure shows the results of removing the first component. (a) The residual image without loop gain from the Asp-Clean algorithm. (b) The residual image with the fixed loop gain $0.7$ from the Asp-Clean algorithm. (c) The residual image with the adaptive loop gain $0 \sim 0.7$ from the Algas-Clean algorithm. These images are displayed in the range from $-0.03$ \textrm{Jy/pixel} to $0.703$ \textrm{Jy/pixel} and scaling power cycles is $-2.0$.}
\label{First-Residual}
\end{figure}

To better illustrate the motivation of our algorithm, we use the 3C31
residual images in Fig~\ref{First-Residual} which are the residuals after removing the
first component to show the effects of loop gain of
  1.0, a fixed loop gain and the adaptive loop gain
respectively. To display clearly, the residual images only show the
region where the first component is removed. From Fig~\ref{First-Residual}(a), we can see
that if the component optimized by the Levenberg-Marquardt
method is removed from the dirty image without a loop gain (that is, 1.0), strong
  negative structures appear in the residual image (as expected), due
  to over-subtraction. This is caused by the fact that a spatial
scale function tries to represent the true sky image as much as
possible by optimizing an objective function on the whole dirty
image. With a fixed loop gain residual structures in Fig~\ref{First-Residual}(b) are much reduced compared to
  the residuals in in Fig~\ref{First-Residual}(a). This demonstrates that the fixed loop
gain can partly suppress over-subtraction. However, the
over-estimation in some parts particularly in the wings of the
  Gaussian component used -- still exists. This reason is that when Gaussian functions are used to represent a component, a
common strategy is that after getting the component, the tails are
truncated at a certain ratio to the peak of the component
\citep{bha04,cor08}. The method is unlikely to be optimal for all
components. For example, when a Gaussian function is used to represent
a flat and finite-support extended structure, the
component derived by fitting a Gaussian to the structure will have extended wings {\it
    beyond} the extent of the true emission. Truncation alone is not good in this
  case and can lead to an improper truncated Gaussian, which can lead
to an over-subtraction in some parts of the image. This is
  evident in Fig~\ref{First-Residual}(b). Using the
adaptive loop gain scheme
smaller loop gains get used to scale-down the wings of the Gaussian component, followed by a truncation . This is more effective than the current method.  It is not
difficult to appreciate the effectiveness after comparing Fig~\ref{First-Residual}(b) with
Fig~\ref{First-Residual}(c). As mentioned earlier, the other major advantage of the
  adaptive loop gain approach is that it not only changes across a given component, it {\it also} varies from one component to the other.

\section{Summary}

Since the first CLEAN algorithm was proposed in
1970s, many variants have been developed. However the fixed loop gain parameter is still used to optimize
components. If we view the reconstruction problem as an approximation
problem, loop gains are actually step lengths of approximation along
gradient directions in an iterative algorithm. They can affect
the imaging performance as well as
  rate of convergence significantly. In this paper, we have devised a new algorithm for a
  shape-dependent adaptive loop gain scheme based on amplitudes of an
  extended component. The adaptive loop gains are designed to solve
  the problems of the representation of finite-support structures and the resulting
  errors in the model image. Tests show that the Algas-Clean algorithm
  can give a more accurate model with fewer components. Fewer
  components also means that the approximation of the true sky image
  with the proposed adaptive loop gain scheme is more effective than
  that with fixed loop gain and also speed up deconvolution
  process. The work is implemented with Python and the CASA package. The script will be available \footnote{ https://github.com/lizhangscience/Algas\_Clean.} soon later and the C++ implementation in the CASA package is underway.

%}
\acknowledgments
% <Acnowledgments text>
We would like to thank S. Bhatnagar and U. Rau for numerous help in
this work. We thank the anonymous referee for helpful comments. This work is supported by the National Basic Research
Program of China (973 program: 2012CB821804 and 2015CB857100), the
National Science Foundation of China (Grant No. 11103055) and the West
Light Foundation of the Chinese Academy of Sciences (Grant
No. RCPY201105).

\clearpage

%% Non-BibTeX  (Name-Year style)
%
% \begin{thebibliography}{}
% \bibitem[\protect\citeauthoryear{<author>}{<year>]{ref:?}
%    <ref. entry>
% \bibitem[\protect\citeauthoryear{<author>}{<year>]{ref:?}
%    <ref. entry>
% \end{thebibliography}

\end{document}